# Rotational control and selective isotope alignment by ultrashort pulses

**Sharly Fleischer, I.Sh. Averbukh and Yehiam Prior**

*Department of Chemical Physics, Weizmann Institute of Science, Rehovot, **Israel** 76100*
yehiam.prior@weizmann.ac.il, phone +972-8-934-4008, fax +972-8-934-4126

**Abstract:** Strong ultrashort laser pulses give rise to prompt molecular alignment followed by full, half and quarter quantum revivals of the rotational wave packets. In Four Wave Mixing (FWM) experiments, we make use of the long succession of repetitive alignment revivals to demonstrate isotope-selective alignment and time-resolved discrimination between different isotopic species. For two isotopes with a commensurate ratio of their moments of inertia, we observe constructive and destructive interferences of the FWM signals originating from different isotopes. Such destructive interference serves as an indication for distinct transient angular distributions of different isotopic species, paving the way to novel time-resolved analytic techniques, as well as to robust methods for isotope separation. When two pulses are used, enhancement or reduction of the degree of rotational excitation and of the corresponding molecular alignment is observed, depending on the delay between pulses. We implement the double pulse excitation scheme in a binary isotopic mixture, enhancing the rotational excitation of one isotopic component while almost totally reducing the rotational excitation of the other.

Alignment and orientation of molecules have always intrigued spectroscopists, and provided a wide range of topics to be studied, from the classical early studies of molecular alignment in constant electric and magnetic fields and the relaxation that follows, to optical activity and polarization effects of oriented molecules in liquid crystals. In the gas phase, molecular alignment following excitation by a strong laser pulse was observed in the seventies [1], and proposed as a tool for optical gating. In the early experiments, picosecond laser pulses were used for the excitation, and deviation of the refractive index from that of an isotropic gas was monitored as evidence for alignment [2,3]. More recently, these observations have been revisited both theoretically and experimentally (for a recent review, see [4]). Spatial and temporal dynamics was studied [5,6,7], and sophisticated pulse sequences were described, giving rise to better alignment [8,9,10]. Further manipulations such as the optical molecular centrifuge and alignment-dependent strong field ionization of molecules were demonstrated [11,12,13]. Molecular phase modulators have been shown to compress ultrashort light pulses [14,15] and molecular alignment has been used for controlling high harmonic generation [16,17,18]. Moreover, experiments based on the femtosecond transient grating technique were reported where ultrafast laser pulses induced molecular alignment and deformation were studied in details [19,20].

In this work we study molecular alignment and rotational wave packets control in isotopic mixtures excited by strong ultrashort laser pulses. The rotational wave packets exhibit tens of full, half and quarter quantum revivals of their alignment before they decay after hundreds of picoseconds. We demonstrate the possibility to obtain substantial alignment of a given isotope at specific times, while keeping the other species of the mixture in a more or less isotropic state. Moreover, we propose and experimentally demonstrate simultaneous alignment of one of the isotopic components and anti-alignment of another one, thus creating a dramatic contrast in their angular distributions. Based on this pronounced effect, a robust scheme for isotope separation is shown to be feasible for rotational wave packets, in a manner similar to the one proposed and demonstrated for vibrational wave packets [21]. Further operations on the isotopic wave packets are possible with the application of additional laser pulses, and we demonstrate a nondestructive selective addressing and manipulation of only one component in a mixture. Since no excited electronic states or specific optical resonances are involved, the present observations pave the way to new generic approaches for isotope ratio analysis and determination.

In our experiments we use a time-delayed degenerate, forward propagating three dimensional phase matched four-wave mixing arrangement[22], where the first two pulses set up a spatial grating of transiently aligned molecules, and the third, delayed pulse is scattered off this grating. In this set of experiments, all three input beams (and therefore the fourth output beam as well) were linearly, vertically polarized. The experiments were carried out with ~100 fs pulses, from a regeneratively amplified Ti:Sapphire laser at 800 nm, and the molecular samples consisted of homonuclear diatomic molecules in the gas phase.

For molecules without a permanent dipole moment ($O_2$, $N_2$, *etc.*), the applied laser electric field induces an electric dipole, which in turn interacts with the electric field. This interaction tends to align the molecules along the direction of the external field and is given by $V \propto -\cos^2\theta$ (where $\theta$ is the angle between the molecular axis and the field polarization). The torque exerted on a molecule by a short, strong laser pulse is $\tau \propto -\sin 2\theta$, which induces an angle-dependent angular velocity distribution in the ensemble. After a short delay, the angular distribution peaks in the direction of the field, i.e. the molecular ensemble exhibits (partial) transient alignment at field-free conditions. This alignment is symmetrical with respect to the direction of the field ("up" or "down") and in what follows we refer to this geometrical shape as "cigar-like". Following the first alignment, the ensemble continues to evolve freely and the aligned state re-appears periodically as a result of the quantum revival effect. The period is referred to as the revival time $T_{rev}=1/2Bc$, where the rotational constant is $B = \hbar/4\pi Ic$, $c$ is the speed of light and $I$ is the molecular moment of inertia. In every cycle, just prior to attaining the cigar shape, the ensemble goes through anti-alignment, namely a state in which the molecules lie in the plane perpendicular to the direction of alignment [9]. In this 'anti-aligned' state, the geometry is of a "disc" shape. In addition to the full revivals described above, the quantum phenomenon of fractional revivals is observed at half, quarter and other rational parts of the revival time, and each such partial revival involves permutations of aligned and anti-aligned shapes. In particular, in the half-revival domain, the order of alignment and anti-alignment is reversed with respect to the full-revival domain, and the cigar-type molecular angular distribution precedes the disc-like one [9]. We note that molecular alignment is reflected in the increase of the gas refractive index, while the anti-alignment causes its reduction compared to the isotropic case.

In the first set of experiments we utilize the periodic behavior of the rotational wave packets to distinguish between molecular isotopes too similar to be resolved on shorter time scale. As a simple example, we consider the homonuclear diatomic chlorine molecule, existing in three different isotopic forms: $^{35}Cl-^{35}Cl$, $^{35}Cl-^{37}Cl$, $^{37}Cl-^{37}Cl$. Due to small differences in their moments of inertia, the isotopic species display very close revival periods. After one half revival period, the signals from the alignment peaks of the three isotopes are overlapping in time, and therefore are unresolved. However, due to the periodicity of the revivals, the difference between revival periods is accumulating and the three isotopic species may be temporally resolved after several revival cycles. In figure 1 a,b , at the first half and full revivals, the signals from the different isotopes are not yet resolved, but at the second full revival (~140 ps for chlorine) (figure.1 a,c) the contributions of the three different isotopic species are separated by about 4 ps and are fully resolved. The relative intensities of the signals are in agreement with the natural abundance of different chlorine isotopes in the mixture. Note that the FWM signal is proportional to the square of the nonlinear induced susceptibility, and therefore only positive signals are observed both for cigar-type and disc-like angular distributions. Based on the above described sequence of alignment and anti-alignment events, the signal from each isotope should have been a simple doubled peak. The observed slight deviations from this simple picture and the weaker 'extra' peaks arise from the centrifugal distortion of higher rotational states.

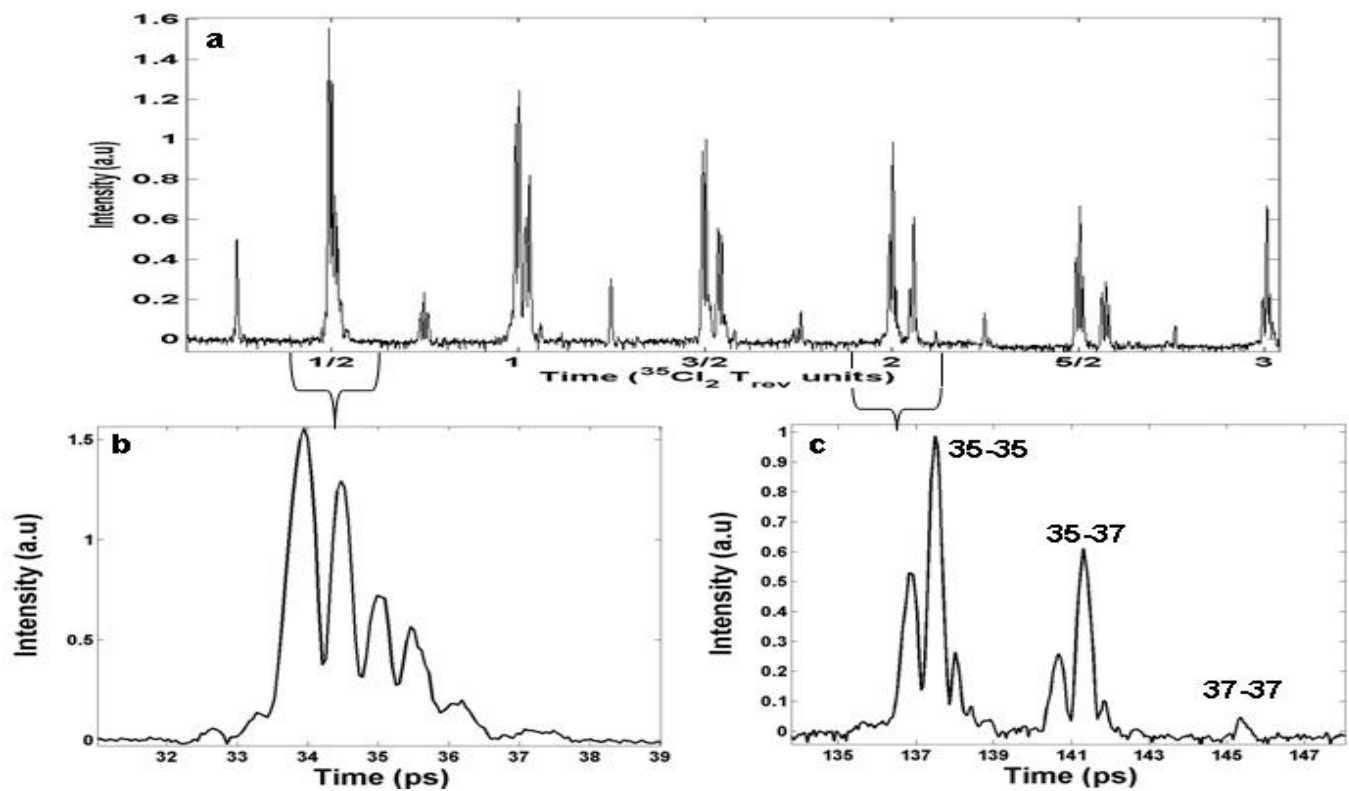

**Figure 1:** a) Long scan FWM signal observed from Cl$_2$ gas (225 torr, room temperature). (b) Enlarged first half revival (~35 ps), the contributions of the different isotopic species temporally overlap. (c) Enlarged second full revival (~140 ps), the fully resolved contributions of the different isotopes are clearly seen.

Next, we report for the first time the simultaneous alignment and anti-alignment of two separate isotopes, observed by interference of their FWM signals. As detailed above, the revival time is inversely proportional (through the rotational constant) to the molecular moment of inertia, which in turn is proportional to the reduced mass. Thus, for a homonuclear diatomic molecule like nitrogen, the revival time ratio for two isotopic species $^{15}N_2$ and $^{14}N_2$ is a rational number (14/15) proportional to their mass ratio. Whenever the alignment (full or fractional revival) of the two isotopes temporally coincide, the contributions of the aligned (cigar) and anti-aligned (disc) states to the nonlinear susceptibility interfere, a fact that can be readily seen in a time delayed four wave mixing experiment. As an example, following an impulse excitation by a short pulse, the 14$^{th}$ full revival of $^{15}N_2$ and 15$^{th}$ full revival of $^{14}N_2$ coincide at ~ 126 ps, and should give rise to constructive interference in the FWM signal.

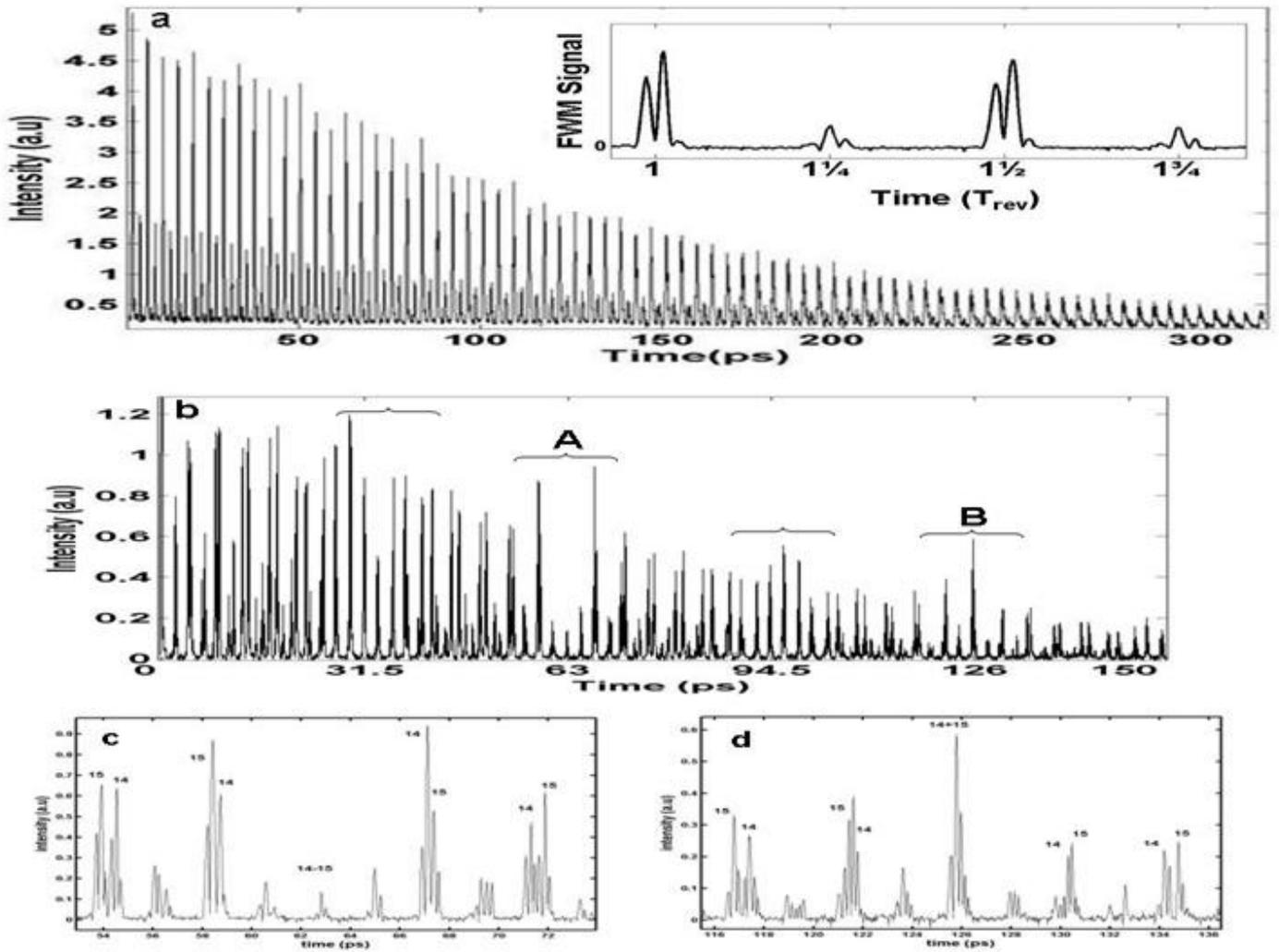

**Figure 2: a)** Long scan FWM signal from $N_2$ gas (200 torr, room temperature). 40 full revivals are shown (~320ps). Quarter, half and full revivals are clearly seen. The inset depicts one full revival cycle. **b)** FWM signal from 1:1 $^{14}N_2 : ^{15}N_2$ mixture (500 torr, room temperature). Destructive interference of full and half revival signals is seen in region A at ~63 ps; Constructive interference of the two full revival signals is observed in region B at ~126 ps. The quarter and half revival signals interfere around 31.5 and 94.5 ps. **c)** Enlarged view of the destructive interference region A. **d)** Enlarged view of the constructive interference region B.

Figure 2a depicts the time delayed degenerate four wave mixing signal obtained from a single isotope of nitrogen ($^{14}N_2$) following strong, ultrashort excitation. Over 40 revival cycles, 8.3 ps each, are observed demonstrating full, half and quarter revivals. The long observed decay time of the revivals (hundreds of ps) originates from collisions within the cell, and time of flight across the beam.

Figure 2b depicts a full scan over many revivals of 1:1 isotopic mixture of $^{14}N_2, ^{15}N_2$. At ~ 63 ps (figure 2b,c), the $^{15}N_2$ isotope completes 7 full revival periods while $^{14}N_2$ performs $7\frac{1}{2}$ of its own revival cycles Due to the reversed order of the alignment and anti-alignment events for these two isotopes, a pronounced destructive interference is observed in the combined FWM signal (see region **A**). The dip at 63 ps presents a clear indication of an extreme angular separation of the isotopic components: when one of the isotopes reaches a cigar state, the other one exhibits a disc-like angular distribution, and vice versa. This provides a favorable configuration for further manipulation such as selective ionization (or dissociation) of the aligned component by an additional linearly polarized laser pulse. At ~126 ps (figure 2b,d), as discussed above, full revivals of the two isotopes coincide, giving rise to constructive interference (region **B**). Other combinations of full and fractional revivals give rise to other interference phenomena (31.5 and 94.5 ps), and these will be discussed in detail in a forthcoming publication.

As the next step, we consider a "nondestructive" control of molecular alignment by a pair of pulses. A short laser pulse always "kicks" the molecules, delivering torque to rotating wave packets. The response of the molecular ensemble, however, is very sensitive to the timing of the second pulse. If a second pulse is applied **at the time of exact revival**, its effect is similar to that of the first pulse, namely it kicks the molecules "in phase" with their rotational motion, and **adds angular momentum** to the already rotating molecules, which results in a more pronounced alignment. This is illustrated by the measurements presented in figure 3a, where the second pulse was applied exactly at $3T_{rev}$.

If, on the other hand, the second pulse is applied **at a time of half revival**, where the molecules are moving away from alignment, the torque impacted by the second pulse effectively **cancels the coordinated motion** of the rotating molecules, thwarting any future revivals. In figure 3b the second pulse was applied at $2.5T_{rev}$. The difference in response is dramatic!

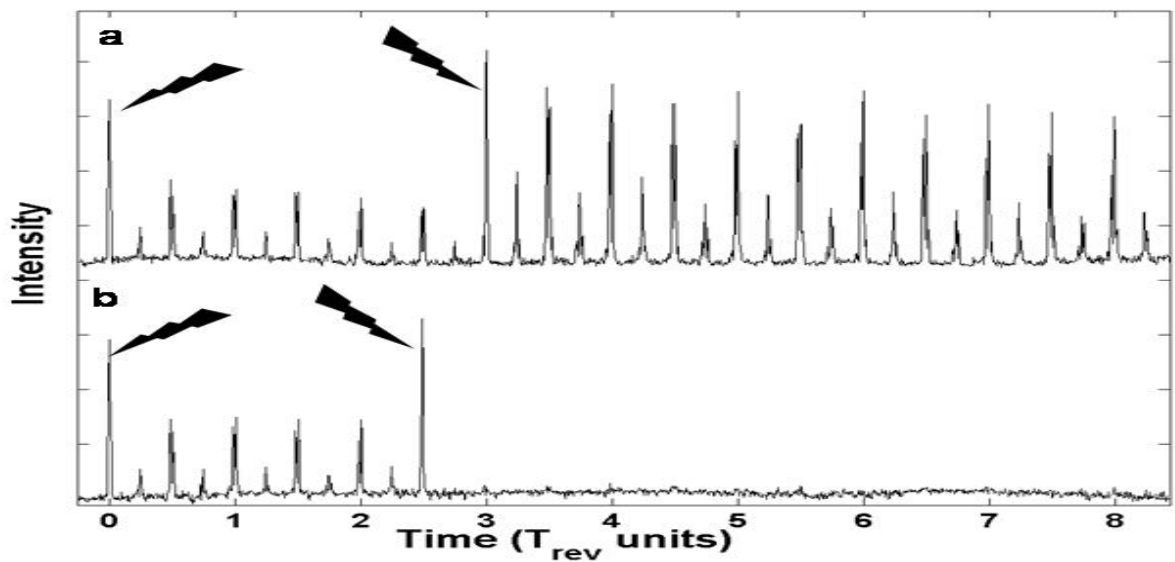

**Figure 3: Alignment signal from $^{15}N_2$ gas (300 torr, room temperature), following excitation by two pulses. a) The two pulses are separated by a multiple of the exact revival time. The torque from the second pulse adds coherently to that from the first one, resulting in enhanced alignment signal. b) The two pulses are separated by an odd multiple of half revival time. The torque from the second pulse is opposite to the molecular angular velocity, resulting in effective stopping of the rotation.**

In the last series of experiments, we used this difference to achieve a two pulse isotope-selective control in the same 1:1 mixture of $^{14}N_2$ and $^{15}N_2$ used above. As detailed before, at ~ 63 ps $^{14}N_2$ completes 7.5 revival cycles while $^{15}N_2$ completes 7 revival periods. In this time domain, one of the isotopes is rotating from the disc plane towards the cigar axis, while the other one goes in the opposite direction. The application of a second pulse at that unique time has a drastically different effect on the two species. As shown in figure 4, the second pulse at ~ 63 ps enhances the alignment of $^{15}N_2$ molecules, and almost completely stops the rotation of $^{14}N_2$ isotopes!

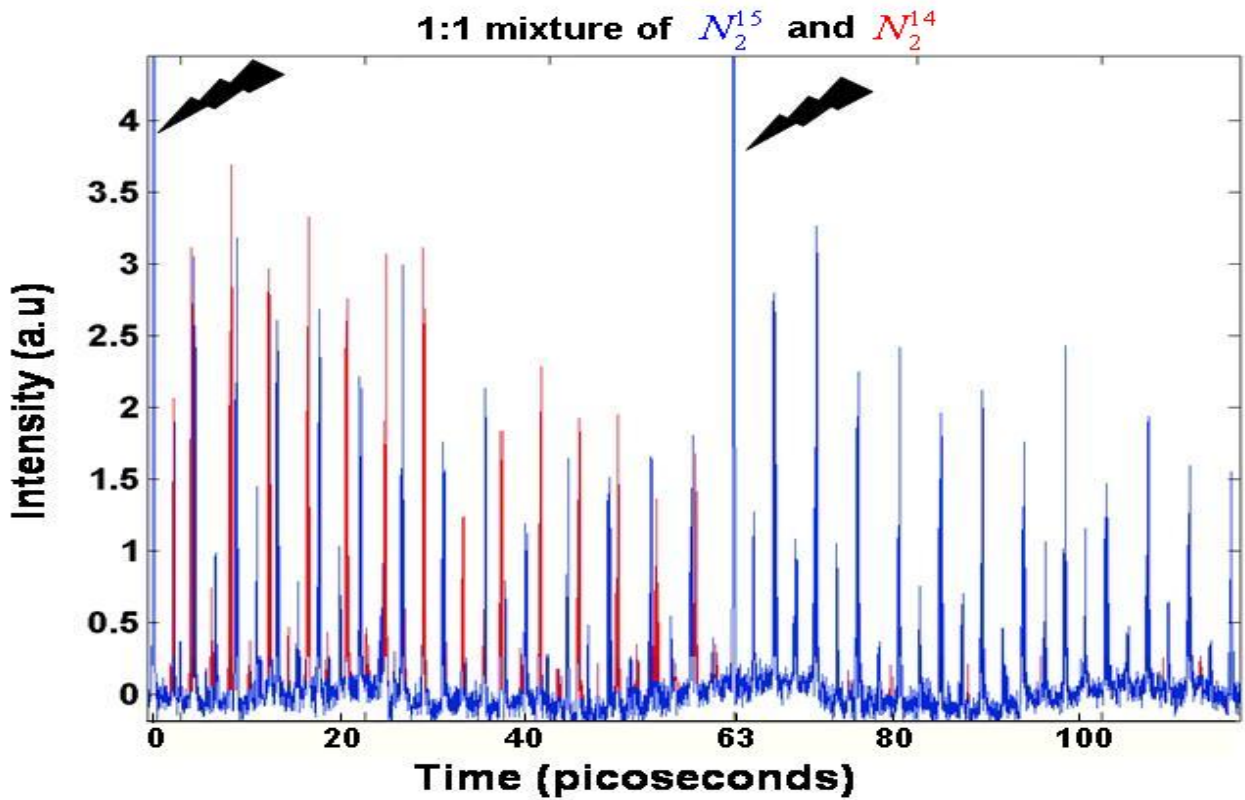

**Figure 4:** FWM signal from 1:1 mixture of $^{14}N_2$, $^{15}N_2$ (500 torr, room temperature) subject to two pulses ('kicks'), delayed by 63ps. The first kick excites both molecular isotopes, but the second kick, affects them in an opposite way. The individual peaks of the different isotopes were artificially colored $^{15}N_2$ - blue and $^{14}N_2$ red. The blue ($^{15}N_2$) signal is enhanced by the second kick while the red ($^{14}N_2$) signal disappears (see text).

In summary, multiple rotational quantum revivals are observed in molecular isotopic mixtures in response to an impulse excitation by a short pulse. By utilizing the repetitive nature of the alignment signal and slight difference in the isotopes revival periods, we have demonstrated time-resolved discrimination between different isotopic components. Moreover, by observing the destructive interference in the FWM signal from an isotopic mixture we have identified the specific times when different isotopic species attain drastically different angular distributions. One isotope becomes field-free aligned, while the other one lies in the plane perpendicular to the alignment direction. Such a configuration is most favorable for isotope-selective ionization or dissociation by an additional laser pulse, which paves the way to an effective and robust new isotope separation technique. When more than one pulse is used for the rotational excitation, the exact timing between pulses is crucial. We demonstrated that if a second pulse is applied at exact full revival, the rotational alignment is enhanced, whereas if the second pulse is properly applied at the half revival time, the rotational periodic alignment may be effectively stopped. We have further showed that this behavior may be implemented in an isotopic mixture, providing a robust methodology of individually addressing a single component in a mixture, and strongly affecting the rotation of selected species in a mixture.

The ability to selectively address a single species in a multi component mixture, and change its physical properties (i.e. alignment, or rotation excitation level) is an important outcome of this work. Based on these observations, one may envisage ultrafast time-resolved analytical methods for isotope ratio determination, identification and discrimination of close chemical species, and trace analysis in multicomponent mixtures. Experiments are under way to utilize some of these new ideas.

We acknowledge the support of the Israel Science Foundation, and the James Franck program at the Weizmann Institute.

[1] M. A. Duguay and J. W. Hansen, Appl. Phys. Lett. **15**, 192 (1969).
[2] C. H. Lin, J. P. Heritage, and T. K. Gustafson, Appl. Phys. Lett. **19**, 397 (1971).
[3] C. H. Lin, J. P. Heritage, and T. K. Gustafson, Appl. Phys. Lett. **34**, 1299 (1975).
[4] H. Stapelfeldt, T. Seideman, Rev. Mod. Phys. **75** (2003) 543.
[5] T. Seideman, J. Chem. Phys. **103**, 7887 _(1995)_.
[6] J. Ortigoso, M. Rodriguez, M. Gupta, and B. Friedrich, J. Chem. Phys. **110**, 3870 _(1999)_.
[7] F. Rosca-Pruna and M. J. J. Vrakking, Phys. Rev. Lett. **87**, 153902 _(2001).
[8] I. Sh. Averbukh and R. Arvieu, Phys. Rev. Lett. **87**, 163601 (2001).
[9] M. Leibscher, I. Sh. Averbukh, and H. Rabitz, Phys. Rev. Lett. **90**, 213001 (2003) ; Phys. Rev. A **69**, 013402 (2004).
[10] M. Renard, E. Hertz, S. Guérin, H. R. Jauslin, B. Lavorel, and O. Faucher, Phys. Rev. A **72**, 025401 (2005)
[11] J. Karczmarek, J.Wright, P.Corkum, and M.Ivanov, Phys. Rev. Lett. **82**, 3420 (1999).
[12] I.V. Litvinyuk, Kevin F. Lee, P.W. Dooley, D.M. Rayner, D.M. Villeneuve, and P.B. Corkum, Phys. Rev. Lett. **90**, 23 (2003).
[13] D. Pinkham and R.R. Jones, Phys. Rev. A **72**, 023418 (2005).
[14] R.A.Bartels, T.C.Weinacht, N.Wagner, M. Baertschy, Chris H. Greene, M.M. Murnane, and H.C. Kapteyn, Phys. Rev. Lett **88**, 019303 (2002).
[15] V. Kalosha, M. Spanner, J. Herrmann, and M. Ivanov, Phys. Rev. Lett., **88**, 103901 (2002).
[16] R. Velotta, N. Hay, M. B. Mason, M. Castillejo, J. P. Marangos, Phys. Rev. Lett. **87**, 183901 (2001).
[17] M. Kaku, K. Masuda, and K. Miyazaki, Jpn. J. Appl. Phys. **43**, L591 (2004).
[18] J. Itatani, D. Zeidler, J. Levesque, M. Spanner, D.M. Villeneuve, P.B. Corkum, Phys. Rev. Lett. **94**, 123902 (2005).
[19] E. J.Brown, Qingguo Zhang and M. Dantus. J.Chem. Phys **110**,5772 (1999).
[20] M. Comstock, V. Senekerimyan, and M. Dantus, J. Phys. Chem. A **107**, 8271 (2003).
[21] I.Sh. Averbukh, M.J.J. Vrakking, D.M. Villeneue and A. Stolow, Phys.Rev.Lett. **77**, 3518 - 3521 (1996).
[22] Y.Prior, App.Opt. **19**, 1741 (1980).